\documentclass[apj]{emulateapj}
\usepackage{apjfonts,mathptmx}

\def\xmm {\emph{XMM-Newton}}
\def\cxo {\emph{Chandra}}
\def\swift {\emph{Swift}}
\def\sax {\emph{BeppoSAX}}
\def\xte {\emph{RossiXTE}}
\def\asca {\emph{ASCA}}
\def\src {SGR\,1627--41}
\def\flux {\mbox{erg cm$^{-2}$ s$^{-1}$}}
\def\lum {\mbox{erg s$^{-1}$}}

\begin{document}

\title{\emph{XMM-Newton} discovery of 2.6 s pulsations in the soft gamma-ray repeater SGR\,1627--41\altaffilmark{*}}

\author{P.~Esposito\altaffilmark{1,2}, A.~Tiengo\altaffilmark{1}, S.~Mereghetti\altaffilmark{1}, G.~L.~Israel\altaffilmark{3}, A.~De~Luca\altaffilmark{1,4}, D.~G\"otz\altaffilmark{5}, N.~Rea\altaffilmark{6}, R.~Turolla\altaffilmark{7,8}, and S.~Zane\altaffilmark{8}}

\altaffiltext{*}{Based on observations obtained with \emph{XMM-Newton}, an ESA science mission with instruments and contributions directly funded by ESA Member States and NASA.}
\affil{$^{1}$ INAF/Istituto di Astrofisica Spaziale e Fisica Cosmica - Milano, via E.~Bassini 15, 20133 Milano, Italy; \url{paoloesp@iasf-milano.inaf.it}}
\affil{$^{2}$ INFN - Istituto Nazionale di Fisica Nucleare, Sezione di Pavia, via A.~Bassi 6, 27100 Pavia, Italy}
\affil{$^{3}$ INAF/Osservatorio Astronomico di Roma, via Frascati 33, 00040 Monteporzio Catone, Italy}
\affil{$^{4}$ IUSS - Istituto Universitario di Studi Superiori, viale Lungo Ticino Sforza, 56, 27100 Pavia, Italy}
\affil{$^{5}$ CEA Saclay, DSM/Irfu/Service d'Astrophysique, Orme des Merisiers, B\^at. 709, 91191 Gif sur Yvette, France}
\affil{$^{6}$ Astronomical Institute ``Anton Pannekoek,'' University of Amsterdam, Kruislaan 403, 1098~SJ Amsterdam, The Netherlands}
\affil{$^{7}$ Dipartimento di Fisica, Universit\`a degli Studi di Padova, via F.~Marzolo 8, 35131 Padova, Italy}
\affil{$^{8}$ Mullard Space Science Laboratory, University College London, Holmbury St. Mary, Dorking, Surrey RH5 6NT, UK}

\shortauthors{P.~Esposito et al.}
\shorttitle{\emph{XMM-Newton} discovery of 2.6 s pulsations in SGR\,1627--41}

\journalinfo{This is an author-created, un-copyedited version of an article accepted for publication in The Astrophysical Journal Letters. IOP Publishing Ltd is not responsible for any errors or omissions in this version of the manuscript or any version derived from it. The definitive publisher authenticated version is available online at 10.1088/0004-637X/111/L1.}
%\journalinfo{This is an author-created, un-copyedited version of an article accepted for publication in The Astrophysical Journal Letters.}
\submitted{Received 2008 October 22; Accepted 2008 November 24}

\begin{abstract}
%After nearly a decade of quiescence, the soft gamma-ray repeater \src\ reactivated on 2008 May 28 with a bursting episode that lasted one day and was accompanied by a flux enhancement. We report on the results of a 120-ks long X-ray observation of the source obtained with the \xmm\ satellite on 2008 September 25--26. We found that the observed 2--10 keV flux was $3.4\times10^{-13}$ \flux\ (well above the pre burst-activation level) and that the spectrum was well fitted by an absorbed power law plus blackbody model (the best-fit parameters are photon index $\Gamma\simeq0.6$, blackbody temperature $kT\simeq0.5$ keV, and absorption $N_{\rm{H}}\approx1.2\times10^{23}$ cm$^{-2}$). We discovered pulsations with period $P=2.594578(6)$ s. The pulse profile is double-peaked, with a pulse fraction in the 2--12 kev energy range of 19\%$\pm$3\% and 24\%$\pm$3\% for the fundamental and the second harmonic, respectively. The spatial analysis reveals a shell of diffuse soft X-ray emission which is likely the first detection at X-rays of the young supernova remnant G337.0--0.1.
After nearly a decade of quiescence, the soft gamma-ray repeater \src\ reactivated on 2008 May 28 with a bursting episode followed by a slowly decaying enhancement of its persistent emission. To search for the still unknown spin period of this SGR taking advantage of its high flux state, we performed on 2008 September 27--28 a 120 ks long X-ray observation with the \xmm\ satellite. Pulsations with $P=2.594578(6)$ s were detected at a higher than 6$\sigma$ confidence level, with a double-peaked pulse profile. The pulsed fraction in the 2--12 keV range is $19\%\pm3\%$ and $24\%\pm3\%$ for the fundamental and the second harmonic, respectively. The observed 2--10 keV flux is $3.4\times10^{-13}$ \flux, still a factor of $\sim$ 5 above the quiescent pre-burst-activation level, and the spectrum is well fitted by an absorbed power law plus blackbody model (photon index $\Gamma\simeq0.6$, blackbody temperature $kT\simeq0.5$ keV, and absorption $N_{\rm{H}}\approx1.2\times10^{23}$ cm$^{-2}$). We also detected a shell of diffuse soft X-ray emission which is likely associated with the young supernova remnant G337.0--0.1.
\end{abstract}

\keywords{ ISM: individual (G337.0--0.1) --- stars: neutron --- supernova remnants --- X-rays: individual (\src) --- X-rays: stars}

\section{Introduction}
%Magnetars are isolated neutron stars with exterior magnetic fields of $10^{14}$--$10^{15}$ G and even stronger fields inside the star, making them the strongest magnets in the Universe \citep{thompson95,thompson96,tlk02}. Two classes of X-ray sources are thought to be ultimately powered by their ultra-strong magnetic fields and thus recognized as magnetars: the anomalous X-ray pulsars (AXPs) and the soft gamma-ray repeaters (SGRs). They 
Two classes of young isolated neutron stars, the anomalous X-ray pulsars (AXPs) and the soft gamma-ray repeaters (SGRs) are currently believed to host magnetars, i.e. isolated neutron stars endowed with ultra-strong magnetic fields, $B\sim 10^{14}$--$10^{15}$ G on the surface and one (or more) order of magnitude higher in the interior \citep{duncan92,thompson93,thompson95}. SGRs and AXPs share a number of properties, including rotation periods in the 2--12 s range, large period derivatives ($\dot P\sim 10^{-13}$--$10^{-9}$ s s$^{-1}$), inferred surface magnetic dipole field strength $B\gtrsim10^{14}$ G, large and variable X-ray luminosities (exceeding that available from the braking of their rotation), and the emission of short bursts (see \citealt{woods06} and \citealt{mereghetti08} for recent reviews).\\
\indent The five known Local Group SGRs were discovered as sources of short ($<$1 s) and intense ($10^{41}$--$10^{43}$ \lum) bursts of gamma rays that they emit during sporadic periods of activity interrupted by long stretches of quiescence.  Once localized through their bursts and flares, all of them were found to be persistent X-ray emitters, with luminosities of about $10^{33}$--$10^{36}$ \lum. In four SGRs, \src\ being the exception,  pulsations were detected with periods from 5 to 8 s and period derivatives of $10^{-11}$--$10^{-9}$ s s$^{-1}$.\\
\indent \src\ was discovered in 1998 June by the \emph{Compton Gamma Ray Observatory}  thanks to the intense bursts it emitted at that time, about 100 in six weeks \citep{woods99}. Soon after the discovery of the bursts, its persistent X-ray emission was detected by \sax\ at  a flux level of $\sim$ $7\times10^{-12}$ \flux\ (unabsorbed, 2--10 keV), corresponding to a luminosity of $\sim$ $10^{35}$ \lum\ for a source distance $d= 11$ kpc \citep{corbel99}. This led to the possible association of the SGR
% a luminosity of $\sim$$10^{35}$ \lum,\footnote{Assuming a distance to the source of 11 kpc \citep{corbel99}.} and the SGR was proposed to be associated 
with the young supernova remnant (SNR) G337.0--0.1 in the CTB\,33 complex \citep{woods99,hurley99_1627}. 
%The quiescent spectrum was well modeled by an absorbed power law ($N_{\rm H}\approx8\times10^{22}$ 
%cm$^{-2}$ and photon index $\Gamma\simeq2.5$; \citealt{woods99}). 
In the following 10 years no further bursting activity was reported while the persistent emission steadily decayed to about $10^{33}$ \lum\ (for $d= 11$ kpc; \citealt{corbel99}), the lowest value ever observed from an SGR. At the same time the spectrum softened \citep{kouveliotou03,mereghetti06}.\\
\indent The long-term fading of \src\ was suddenly interrupted by its burst reactivation in 2008 May. This episode was associated with a temporary enhancement of the persistent X-ray flux and a marked spectral hardening \citep{eiz08}. \\
%No further bursting activity has been reported since then.\\
\indent The detection of SGR-like bursts from \src\ made it a bona fide member of the SGR class. However, two strong pieces of evidence in favor of this identification, the measure of the source period and period derivative, were still missing.\footnote[9]{\citet{woods99} reported a candidate periodicity of marginal significance at 6.41 s, but the signal was not confirmed by subsequent observations.}
%Although the membership of \src\ in the magnetar class, testified by its SGR-like bursting activity, has never been questioned, the detection of a spin-period, which is a key parameter for magnetars, was missing yet. 
In order to search for pulsations taking advantage of the enhanced flux after the reactivation, we requested a Target of Opportunity \xmm\ observation, that was performed as soon as the satellite visibility constraints allowed it.
%to be performed as soon as it became visible, in order to deeply search for pulsations taking advantage of the enhanced flux. The one-orbit-long observation was performed on 2008 September 25. In this Letter we report on the discovery of pulsations and other results from this observation.

\section{Observation, data analysis, and results}
Our deep \xmm\ observation of \src\ 
%(ID: 0560180401) 
started on 2008 September 25 at 00:51:02 UT and lasted about 120 ks,
% to September 26 11:31:35 UT, 
occupying the whole satellite revolution number 1611. The data were collected with the EPIC instrument, which consists of two MOS \citep{turner01} and one pn \citep{struder01} CCD cameras sensitive to photons with energies between 0.1 and 15 keV. The EPIC pn was operated in full frame mode (time resolution 73 ms) while both MOS units were in small window mode (time resolution 0.3 s). All the detectors mounted the thick optical blocking filter.\\
\indent The raw observation data files (ODF) were processed with the \xmm\ Science Analysis Software\footnote[10]{See \url{http://xmm.esac.esa.int/external/xmm\_data\_analysis/}} (SAS version 8.0.0) and the calibration files released in 2007 August using standard pipeline tasks (epproc and emproc for the pn and MOS, respectively). The source photons for the timing and spectral analysis were accumulated from a circular region ($30$ arcsec radius) centered on \src. The background events were extracted from source-free regions of the same chip as the source.
%: annular regions with radii of \ldots $\arcsec$ and \ldots $\arcsec$ for the MOS, and a rectangular region with area of $\ldots\times10^3$ arcsec$^2$ located on the side of the source for the pn. 
We selected events with pattern 0--4 and pattern 0--12 for the pn and the MOS, respectively. For the timing analysis, the photon arrival times were converted to the solar system barycenter using the SAS task barycen. Photons having energies below 2 keV and above 12 keV were ignored, owing to the very few counts from \src. A total of about $2650\pm60$ counts above the background were collected from \src\ by the pn detector in the 2--12 keV range, $760\pm40$ by the MOS\,1 detector, and $900\pm40$ by the MOS\,2 detector.

\subsection{Timing analysis}
\begin{figure}
\resizebox{\hsize}{!}{\includegraphics[angle=0]{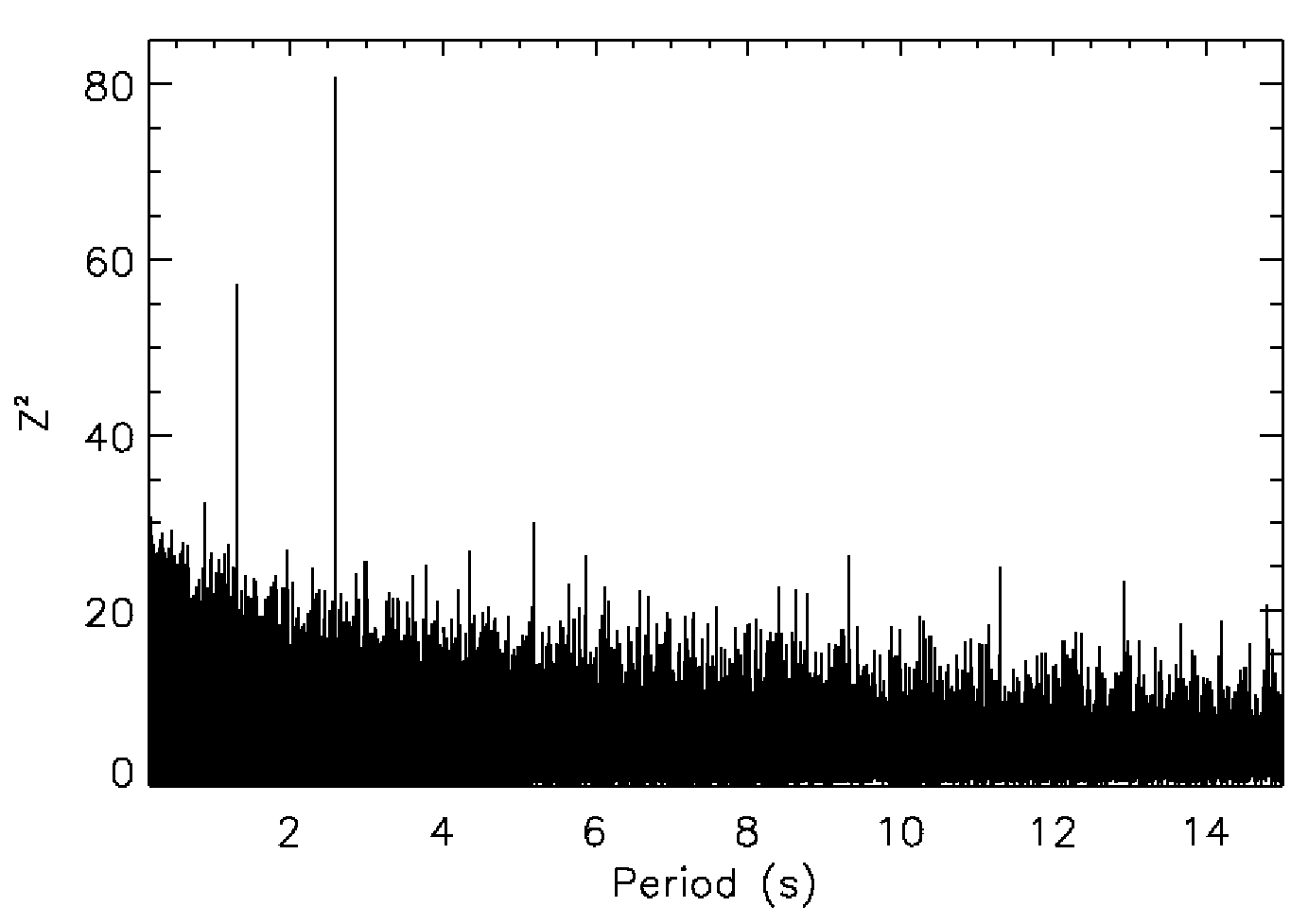}}
\caption{\label{zsquare}$Z^2_2$ periodogram computed for the pn data of \src. The 2.594578s pulsed signal and its first harmonic stand out well above the noise level.}
\end{figure}

\begin{figure}
\centering
\resizebox{0.7\hsize}{!}{\includegraphics[angle=-90]{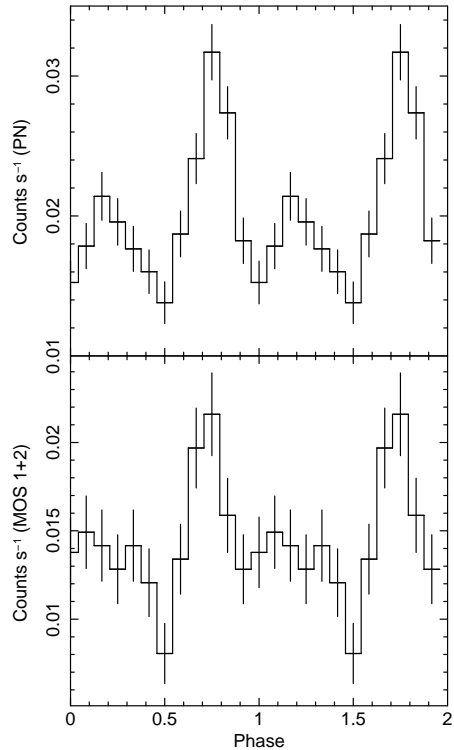}}
\caption{\label{efold} Epoch-folded pulse profile of \src\ (2--12 keV) for the pn (\emph{top}) and MOS data (\emph{bottom}). The arrival time of the pulse
minima (arbitrarely adopted as phase 0) was 54734.0000000(4) MJD (TDB).}
\end{figure}

For the timing analysis we started with the high time-resolution pn data. We searched for pulsations in the 0.15--15 s range\footnote[11]{The frame time of the pn in full frame mode implies a Nyquist limit of about 146.6 ms.} using the $Z^2_n$ test \citep{buccheri83}, with the number of harmonics $n$ being varied from 1 to 5. The most significant peak occurred at about 2.6 s for $n=2$ (see Figure \ref{zsquare}). Taking into account the number of searched harmonics and periods (1,647,257), the $Z^2_2$ value of 80.72 for this peak corresponds to a chance probability of $\sim$ $9\times10^{-10}$ (that is a $>$6$\sigma$ detection).
%%%%%%%%%%%%%%%%%%%FIGURE 1
\\
%%%%%%%%%%%%%%%%%%%FIGURE 1
% \indent  To better estimate the period, we used the epoch folding technique and fitted the peak in the $\chi^2$ versus trial period distribution as described in \citet{leahy87} obtaining $P=2.594578(6)$ s (1$\,\sigma$ uncertainties in the last significant digit are quoted in parenthesis). The corresponding pn and MOS folded lightcurves are shown in Figure \ref{efold}.\\,
\indent We refined the period measurement by folding the pn plus MOS data at the best period inferred through the $Z^2_2$ test and studying the phase evolution by means of a phase-fitting technique (see  \citealt{dallosso03} for more details). The resulting best-fit period was $P=2.594578(6)$ s (epoch 54734.0 MJD; hereafter all uncertainties are at 1$\sigma$ confidence level, unless otherwise specified). The current data set a 3$\sigma$ upper limit on the period derivative of $|\dot{P}|<6\times 10^{-10}$ s s$^{-1}$.\\
%   No meaningful constraints on the period derivative could be derived from these \xmm\ data. \\
%\indent The background-subtracted lightcurves folded at the best period are shown in Figure \ref{efold}. The pulse profile is double-peaked, with a pulsed fraction (semi-amplitude of modulation divided by the mean source count rate) of $19\%\pm3\%$ and $24\%\pm3\%$ for the fundamental and the second harmonic, respectively.
\indent The background-subtracted lightcurves folded at the best period are shown in Figure \ref{efold}. The pulse profile is double peaked, with a pulsed fraction (semiamplitude of sinusoidal modulation divided by the mean source count rate; see \citealt{israel96} for more details) of $19\%\pm3\%$ and $24\%\pm3\%$ for the fundamental and the second harmonic, respectively (the above values correspond to $16\%\pm4\%$ in terms of root mean square, consistently with the 10\% upper limit reported in \citealt{woods06}).
%%%%%%%%%%%%%%%%%%%FIGURE 2
%%%%%%%%%%%%%%%%%%%FIGURE 2
Within the statistical uncertainties the pulsed fraction is energy independent and no significant pulse shape variations as a function of energy were found by dividing the counts in soft and hard energy intervals. To assess this, we compared the folded lightcurves in various energy ranges using a two-dimensional Kolmogorov--Smirnov test \citep{peacock83,fasano87}. The results show that the probability that they come from the same underlying distribution is always larger than 30\%.\\
\indent Knowing the spin period, we went back to archival data and looked for pulsations with enhanced sensitivity. However, our searches in the previous \xmm\ observations, as well as in \sax, \asca\ and \xte\ data,\footnote[12]{Public available \cxo\ data of \src\ could not be used owing to their inadequate frame times.} were inconclusive. The most constraining limits on the pulsed fraction were obtained from \sax\ (1998 August) and \asca\ (1999 February) archival observations of \src. A Fourier analysis of the background-subtracted light curves was performed by means of the method described in \citet{israel96} and no significant periodicity was found in the period interval 2.4--2.8 s (identified by considering the 3$\sigma$ upper limit on $\dot{P}$ inferred above). The 3$\sigma$ upper limits on the pulsed fraction (as defined above), computed according to \citet{vaughan94}, are about 63\% and 51\% for the \sax\ and \asca\ data sets, respectively.\footnote[13]{These limits do not take into account the contribute from the diffuse structures described in Section~\ref{imaging}, that are not resolved in the \sax\ and \asca\ data.}

\subsection{Spectral analysis}
\begin{figure*}
\centering
\resizebox{.6\hsize}{!}{\includegraphics[angle=-90]{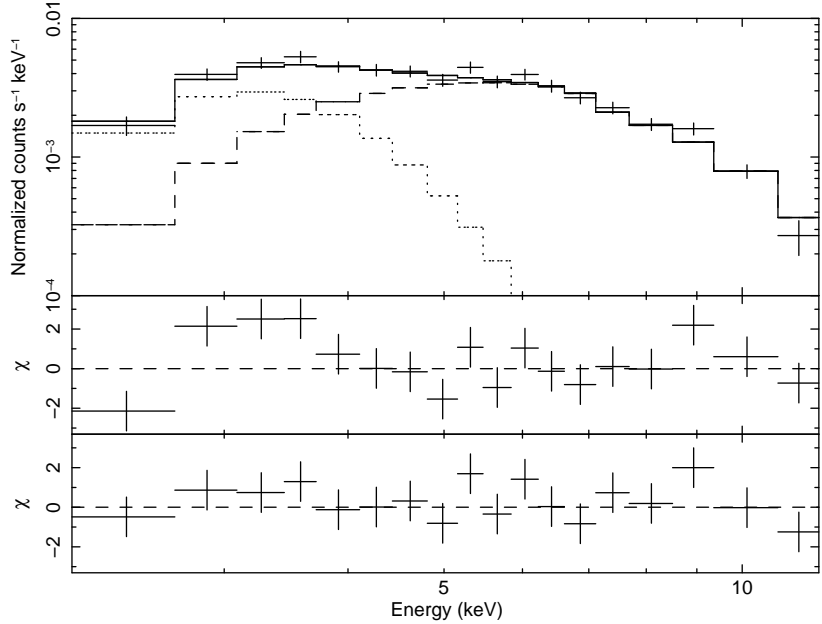}}
\caption{\label{spec} EPIC/pn spectrum of \src. Top: data and best-fit power law plus blackbody model (solid line); the dotted (blackbody) and dashed line (power law) show the individual components. Middle: residuals from the power-law best-fit model in units of standard deviation. Bottom: residuals from the power law plus blackbody best-fit model in units of standard deviation. The data have been rebinned in the plot to better visualize the trend in the spectral residuals.}
\end{figure*}

For the spectral fitting (with XSPEC version 12.4; \citealt{arnaud96}), data were grouped so as to have at least 30 counts per energy bin. The ancillary response files and the spectral redistribution matrices were generated with the SAS tasks arfgen and rmfgen, respectively. We jointly fitted the spectra by pn, MOS\,1, and MOS\,2 to blackbody, power law, and blackbody plus power law models, all corrected for interstellar absorption (see Table~\ref{spectable} for the best-fit model parameters).\\
\indent The data disfavor the blackbody model ($\chi^2_r=1.38$ for 152 degrees of freedom (dof)) and also the power-law fit yields a relatively high $\chi^2$ value ($\chi^2_r=1.19$, 152 dof) with structured residuals (see Figure \ref{spec}). A better fit is obtained by a power law plus blackbody model ($\chi^2_r=1.07$, 150 dof). The best-fit parameters are photon index $\Gamma\simeq0.6$, blackbody temperature $kT\simeq0.5$ keV, and absorption $N_{\rm{H}}\approx1.2\times10^{23}$ cm$^{-2}$. For a distance of 11 kpc \citep{corbel99} this corresponds to a luminosity of $\sim$ $10^{34}$ \lum\ (2--10 keV, unabsorbed).
%%%%%%%%%%%%%%%%%%TABLE
\begin{deluxetable}{lccc}
% \tabletypesize{\scriptsize}
  \tablecolumns{2}
\tablewidth{0pt}
\tablecaption{Spectral results for \src. \label{spectable}}
\tablehead{
\colhead{Parameter} & \colhead{} & \colhead{Value}& \colhead{}\\
\cline{2-4}
\colhead{} & \colhead{BB} & \colhead{PL}& \colhead{PL+BB}
}

\startdata
$N_{\rm{H}}$ ($10^{22}$ cm$^{-2}$) & $1.7\pm0.4$ & $5.2^{+0.6}_{-0.5}$ & $12\pm2$\\
$\Gamma$ & \nodata & $0.94^{+0.06}_{-0.08}$ & $0.6^{+0.1}_{-0.2}$ \\
$kT$ (keV) & $2.5^{+0.2}_{-0.1}$ & \nodata & $0.50^{+0.04}_{-0.06}$  \\
$R$ (km) & $0.041\pm0.003$ & \nodata & $1.2\pm0.1$\\
$\chi^2_r$/dof & 1.38/152 & 1.19/152 & 1.07/150\\
Flux\tablenotemark{a} ($10^{-13}$ \flux) & 3.3 & 3.3 & \phantom{1.0}3.4
\enddata
\tablecomments{The abundances used are those of \citet{anders89} and photoelectric absorption cross-sections from \citet{balucinska92}. Blackbody temperature and radius are calculated at infinity and assuming a distance to the source of 11 kpc \citep{corbel99}.}
%\tablenotetext{a}{Calculated at infinity and assuming a distance to the source of 11 kpc \citep{corbel99}.} 
\tablenotetext{a}{Observed flux in the 2--10 keV range.}
\end{deluxetable}
%%%%%%%%%%%%%%%%%%%TABLE
%%%%%%%%%%%%%%%%%%%FIGURE 3
\\
%%%%%%%%%%%%%%%%%%%FIGURE 3
\indent We compared the current spectral shape of \src\ with those inferred from the latest observations by means of simultaneous fits. Apart from an overall normalization factor (to account for the different luminosities), the current spectrum is consistent with that observed by \swift/XRT during the onset of the 2008 May outburst, while it is significantly harder than that observed by \xmm\ before the outburst, in 2008 February \citep{eiz08}.

\subsection{Imaging analysis}\label{imaging}
\begin{figure*}
\centering
\resizebox{.6\hsize}{!}{\includegraphics[angle=0]{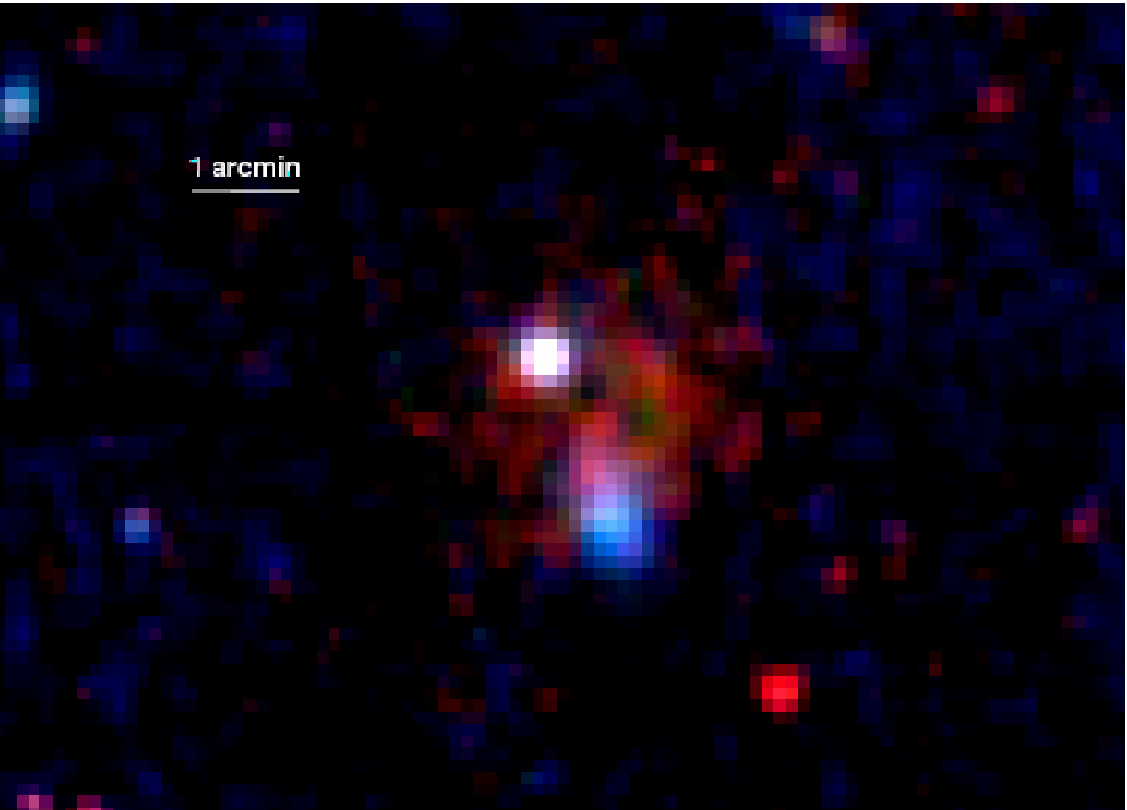}}
\caption{\label{field}EPIC/pn image of the field of \src\ ($11\arcmin\times7\arcmin$). Data from the long 2008 September observation are combined with previous \xmm\ data, totaling to 147.8 ks exposure time. Photon energy is color-coded: red corresponds to the 1.7--3.1 keV energy range, green to 3.1--5.0 keV, blue to 5.0--8.0 keV. North is up, east is left. \src, in white, is the brightest source in the field. The image unveils diffuse X-ray emission from SNR G337.0--0.1, prominent in the softest energy range. A patch of hard diffuse emission is also apparent, southwest of the SNR, possibly due to a background, very absorbed cluster of galaxies.}
\end{figure*}

The energy-coded image of the field around \src\ (Figure \ref{field}) shows that the source is embedded in a complex region of diffuse emission. A shell-shaped structure ($\sim$ 2 arcmin diameter) with a softer spectrum than that of the SGR, as well as a bright spot of remarkably hard diffuse emission ($\sim$ 2 arcmin to the southwest of \src) are apparent in Figure \ref{field}, where red and blue correspond to soft and hard X-ray photons, respectively. The combination of large absorption, source confusion and small photon statistics makes spectral analysis of such diffuse structures rather difficult. Selecting photons from a $\sim$ $100\arcsec\times50\arcsec$ rectangular region located between \src\ and the hard spot, we observe the emission of the very faint diffuse shell to peak between 2 and 3 keV. 
%(where background contamination is of order $\sim50\%$). 
Using \cxo\ imaging data of the field, we estimated the contribution of unresolved point-like sources to the diffuse shell to be of order 30\% in the 1.7--3.1 keV energy range. The brightest point sources in the region are also resolved in  the EPIC data below 1.7 keV, where the diffuse emission is not seen. The spectrum of the shell is adequately described ($\chi^2_r=1.13$, 106 dof) by an absorbed ($N_{\rm{H}} \sim 8\times10^{22}$ cm$^{-2}$) hot plasma model (pshock in XSPEC, $kT\sim1$ keV), with clear indications for emission lines from S (at 2.46 and 3.10 keV) and possibly from Fe (at 6.4 keV) and is therefore consistent with emission from a very absorbed SNR. The net (point source subtracted) surface brightness of the structure is of $\sim$ $10^{-14}$ erg cm$^{-2}$ s$^{-1}$ arcmin$^{-2}$ in the 1.7--3.1 keV band.
%%%%%%%%%%%%%%%%%%%FIGURE 4
\\
%%%%%%%%%%%%%%%%%%%FIGURE 4
\indent The spectrum of the second, harder, diffuse structure was extracted using a 20 arcsec radius circle around the apparent centroid of the feature. The source emission emerges from background above $\sim$ 2 keV and peaks around 5 keV. Significant signal is detected up to 12 keV. A thermal plasma model (mekal in XSPEC, abundance parameter left free) yields an acceptable description of the data ($\chi^2_r=1.27$, 53 dof), although the absorbing column and the temperature are not well constrained ($N_{\rm{H}} > 1.3\times10^{23}$ cm$^{-2}$, $kT>13$ keV at 90\% confidence level for a single parameter). The possible presence of an emission feature at $\sim$ 6.2 keV makes the fit improve after freeing the redshift parameter ($\chi^2_r=1.08$, 52 dof; best-fit $z\sim0.1$, pointing to Fe emission at 6.9 keV in the source frame). The 2--10 keV flux of the source within the aperture is of $\sim$ $1.2\times10^{-13}$ \flux. A possible interpretation is that this source is a background cluster of galaxies. We note that a simple absorbed ($N_{\rm{H}} \sim 1.4\times10^{23}$ cm$^{-2}$) power law ($\Gamma\sim1.6$) also yields an acceptable fit ($\chi^2_r=1.27$, 53 dof). Thus, the source could also be a pulsar wind nebula, unrelated to \src. 

\section{Discussion and conclusions}
 We have discovered pulsations with period of 2.6 s in \src, which was the only known magnetar with no spin period yet measured. The pulse shape is double peaked, with pulsed fractions of $\sim$ $19\%$ and $24\%$ in the two harmonics. No dependence of the pulse profile on the energy band has been found. Among the known magnetars, only 1E\,1547.0--5408 has a shorter period (2.07 s; \citealt{camilo07}) than that of \src.
% No meaningful constraints on the period derivative can be derived from this observation. 
We searched for a similar periodicity in all available archival data with sufficient time resolution, but no pulsation was found. Relying only on the last \xmm\ observation we could set an upper limit on the period derivative,  $|\dot{P}|<6\times 10^{-10}$ s s$^{-1}$ which is not particularly constraining for a magnetar, and yields $B\lesssim1.2\times10^{15}$ G. Based on the peak luminosities of bursts observed from \src, its dipole magnetic field was estimated to be $B\gtrsim1.8\times10^{14}$ G \citep{woods99,eiz08}. This limit can be used to infer a period derivative $\dot{P}\gtrsim1.2\times10^{-11}$ s s$^{-1}$, a spin-down luminosity $\dot{E}\gtrsim8\times10^{33}$ \lum, and a characteristic age $\tau_c= P/(2\dot{P})\lesssim3.4$ kyr. Overall, and lacking any information on the pulsation period and profile in the past, we can only conclude that the timing properties of \src\ reported here appear quite typical of a member of the SGR class.\\
%\indent The timing properties of SGRs are know to show evolution both in time and energy. The lightcurve of SGR\,1806--20 in the pre-giant flare epoch was single-peaked and the (energy-dependent) pulsed fraction in the range $\sim$$6\%$--$14\%$ \citep{mte05}. After the giant flare the pulsed fraction decreased to $\sim$$3\%$ \citep{rea05} then raised again to $\sim$$11\%$ in 2006 September \citep{emt07}. Earlier observations (1996--2001) showed that the source may exhibit more complex pulse shapes \citep{gogus02}. The pulse profile of SGR\,1900+14, as derived from two \xmm\ observations in 2005 September and 2006 April during a period in which the source was quiescent, is single peaked and the pulse fraction ranges from $\sim$$16\%$ to $20\%$ in the different energy bands \citep{met06}. In the past the pulse shape showed rather sharp changes mostly related to the occurrence of dramatic events (the giant flare of 1998 August 27 and also the intermediate flare of 2001 April 18), typically evolving from a more complex (multi-peaked) to a simpler (more sinusoidal) pattern after the flare \citep[see again][]{gogus02}. \cxo\ observations of the quiescent counterpart of SGR\,0526--66 revealed a pulsed fraction of $\sim$$10\%$ \citep{kulkarni03}.\\
\indent Our observation  caught \src\ at a flux of $3.4\times10^{-13}$ \flux, the highest seen with \xmm\ and a factor of $\sim$ 5 above the level preceding its 2008 May burst activation (see Figure~1 of \citealt{eiz08} for a complete flux history of the source). Its spectrum was well fitted in earlier data by a single-component model, either a heavily absorbed power law or a blackbody \citep{mereghetti06}. The higher-quality spectrum of the new data cannot be adequately fitted by these simple models, while it is consistent with a power law plus blackbody model, and it is significantly harder than in any observation of \src\ taken before 2008 May \citep{kouveliotou03,mereghetti06}. The presence of a soft thermal component and the spectral hardening as the source moves from a quiescent to a (burst) active phase makes \src\ akin to other SGRs and AXPs (e.g. \citealt{mte05,rea05}). This latter behavior can be interpreted in the framework of the ``twisted-magnetosphere model'' as due to a progressive growth of the shear in the magnetosphere as magnetic helicity is transferred from the internal to the external field \citep{tlk02,fernandez07,nobili08}.\\
\indent \src\ has been proposed to be associated (\citealt{woods99,hurley99_1627}, but see also \citealt{gaensler01}) to the radio SNR G337.0--0.1 \citep{frail96,whiteoak96}. The shell of diffuse soft radiation (Section~\ref{imaging}), consistent with emission from a young SNR at the distance of \src, is likely to be the first detection of this SNR in X-rays. Due to the complexity of the region, the extension of the SNR at radio frequencies is debated: \citet{sarma97} claimed a diameter of $\sim$ 95 arcsec with \src\ outside the shell, while \citet{brogan00} proposed a different morphology extending the SNR size towards the southwest and encompassing the SGR. The position and  extension of the diffuse X-ray emission are consistent with the latter hypothesis, supporting the possible association between \src\ and G337.0--0.1.\\

\acknowledgements
We thank Norbert Schartel and the staff of the \xmm\ Science Operation Center for performing this Target of Opportunity observation. We thank Pat Romano, Neil Gehrels and Rhaana Starling for their help in planning the observation. We also thank Fabio Gastaldello and Marta Burgay for their helpful advice on the analysis of the field of \src. The Italian authors acknowledge the partial support from ASI (ASI/INAF contracts I/088/06/0 and AAE~TH-058). D.G. acknowledges the CNES for financial support. N.R. is supported by an NWO Veni Fellowship. S.Z. acknowledges support from STFC.

\end{document}